\documentclass{article}

\usepackage{arxiv}

\usepackage{textcomp}
\usepackage{amsmath}
\usepackage{graphicx}
\usepackage{multirow}
\usepackage[colorlinks=true]{hyperref}
\usepackage{subcaption}
\usepackage{svg}
\usepackage{mathtools}
\usepackage{commath}
\usepackage{gensymb}
\usepackage{array}
\usepackage{siunitx}
\usepackage[sorting=none]{biblatex}
\usepackage{booktabs}
\usepackage{ragged2e}

\newcommand{\msSI}[1]{\SI{#1}{\milli\second}}
\newcommand{\orcid}[1]{\href{https://orcid.org/#1}{\includegraphics[width=8pt]{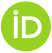}}}
\newcommand{\rev}[1]{\textcolor{black}{#1}}
\newcommand{\revv}[1]{\textcolor{black}{#1}}

\newcommand{\ack}[1]{
\section*{Acknowledgments}
\if@anonymous Removed for anonymity \else #1 \fi}

\newcommand{\funding}[1]{
\section*{Funding}
\if@anonymous Removed for anonymity \else #1 \fi}

\newcommand{\data}[1]{
\section*{Data availability}
\if@anonymous Removed for anonymity \else #1 \fi}

\newcommand{\roles}[1]{
\section*{Author contributions}
\if@anonymous Removed for anonymity \else #1 \fi}

\title{Forecasting the first Edge Localized Mode (ELM) after LH-transition with a neural network trained on Doppler Backscattering data from DIII-D.}

\author{ N.Q.X Teo$^{1,2}$\orcid{0009-0001-6816-6766}\thanks{Future Energy Acceleration \& Translation (FEAT), Strategic Research \& Translational Thrust (SRTT), Agency for Science, Technology and Research (A*STAR), Singapore, Singapore} \\
	\texttt{nqteo@ualberta.ca} 
	\And
	K. Barada$^{3}$\orcid{0000-0001-7724-8491} 
	\And
	V.H. Hall-Chen$^{1,2}$\orcid{0000-0001-6009-3649}
	\And
	L. Gu$^{4}$\orcid{0000-0002-7419-6240} 
	\And
	T.L. Rhodes$^{3}$\orcid{0000-0002-8311-4892} \\ \\
    $^1$Future Energy Acceleration \& Translation (FEAT), Strategic Research \& Translational Thrust (SRTT)\\ Agency for Science, Technology and Research (A*STAR), Singapore, Singapore \\
    $^2$School of Physical and Mathematical Sciences, Nanyang Technological University, Singapore, Singapore \\
    $^3$Department of Physics and Astronomy, University of California, Los Angeles, CA, USA 90095 \\
    $^4$Tohoku University, 41 Kawauchi, Aoba, 980-8576, Sendai, Japan \\
    }

\begin{document}

\maketitle








\begin{abstract}
In H-mode tokamak and stellarator plasmas, edge localized modes (ELMs) lead to the expulsion of heat and particles beyond the edge transport barrier. ELMs cause a loss of energy and have the potential to damage the divertor and other plasma facing components, which motivates efforts to forecast such events to work alongside mitigation systems. In this paper, we use the Doppler backscattering (DBS) diagnostic data as input to train a neural network model, adapted from DeepHit [Lee et al., \textit{Deephit}, AAAI 2018], to forecast the first ELM crash of H-mode discharges in DIII-D. The model takes \msSI{50} of DBS spectrogram data and predicts the probability of an ELM crash occurring within set time windows. Training and testing on shots found in the DIII-D database, we find the initial results promising, with the model reliably forecasting the first ELM \msSI{100} before it occurs. This successful proof-of-concept lays a strong foundation for a predictive tool that can deploy ELM-mitigation techniques before an ELM crash occurs. Future work will expand the training set with carefully selected shots and refine the neural network architecture to improve model robustness to noise and data variation.
\end{abstract}

\section{Introduction}
Edge-localized modes (ELM) are events that occur in H-mode operation of tokamaks and stellarators where the edge plasma is expelled beyond the edge transport barrier~\cite{zohm_edge_1996, garcia2000edge}, resulting in a loss of energy and particles. The ELM events have the potential to damage the divertor and other plasma facing components. In ITER~\cite{shimada2007overview}, large type-I ELMs with energy losses exceeding \SI{1}{MJ}~\cite{maingi2014enhanced} are projected to impose peak loads that can severely damage plasma-facing components~\cite{eich2017elm,loarte2003elm}. \rev{Moreover, in a fusion power plant, even ELMs too small to cause acute melting can burn through divertor detachment, exposing the targets to unacceptable sputtering and erosion. \revv{Consequently,} type-I ELMs will need to be avoided entirely in devices such as ARC~\cite{eich2017elm}. Since a power plant will also be far more sparsely diagnosed than ITER, ELM forecasting there must rely on the few diagnostics, such as Doppler backscattering (DBS), compatible with the burning-plasma environment.} To mitigate this risk, ELM suppression is a critical requirement. \rev{One of the main approaches relies on resonant magnetic perturbations (RMPs)~\cite{evans2008rmp, nazikian2015pedestal}, which have demonstrated varying degrees of success, particularly in plasmas with lower collisionality, higher q$_{95}$ (the safety factor at the flux-surface that encloses the 95\% of the poloidal magnetic flux), and stronger toroidal rotation. However, it is difficult to attain ELM suppression at higher densities~\cite{paz2024plasma}. Other important techniques that have been applied are ELM pacing using pellets~\cite{lang2004elm, baylor2009pellet} or vertical jogs~\cite{gerhardt2010first, de2016understanding}, impurity or Lithium injection~\cite{maingi2014enhanced, osborne2015enhanced}, and operating at small-ELM or no-ELM regimes~\cite{viezzer2018access, paz2021plasma}.} Forecasting the onset of the first ELM using machine learning (ML) provides a valuable complementary tool. By enabling real-time forecasting of the imminent ELM crash, such models can allow early triggering of mitigation actuators or adaptive control of plasma conditions, thereby reducing the risk of large unmitigated ELMs. In addition, ML-based forecasting can help identify subtle precursors in experimental diagnostics that are difficult to detect with conventional methods, potentially improving the reliability and responsiveness of ELM control strategies.

\revv{Although} not traditionally the primary diagnostic for studying ELMs, ELM signatures have been observed by the DBS diagnostic~\cite{burrell2016discovery, ponomarenko2023investigation, yashin2023determination, teo2024using, yashin2026experimental}. The DBS diagnostic injects microwave radiation into the plasma and measures the corresponding backscattered power, which is in turn proportional to the square of turbulent density fluctuations' amplitudes~\cite{hirsch_doppler_2001, hall2022beam, shi20262d}. Due to strong refraction of the probe beam near the cutoff layer, the electric field in that region is large, resulting in the received backscattered signal being dominated by the area near the cutoff~\cite{hirsch_doppler_2001, ruiz2025beam, conway2025doppler, shi20262d}. Additionally, the Doppler shift of the backscattered signal enables background plasma flows to be measured~\cite{hirsch_doppler_2001, pratt2022comparison}. As such, DBS diagnostics have been installed in many tokamaks and stellarators worldwide~\cite{pratt2023density, chowdhury2023novel, macwan2024new, shi2025measurement, conway2025plasma, rienacker2025survey}. DBS has the advantage of having high temporal resolution ~\cite{peebles2010novel}, remote operation, no dependency on neutral beam injection, and greater robustness to damage~\cite{volpe_prospects_2017} compared to more traditional ELM diagnostics, such as photodiodes measuring divertor Deuterium-alpha (D$_\alpha$) line diagnostics~\cite{heidbrink_hydrogenic_2004}, which may be beneficial for the harsh environment of future burning plasmas. \revv{This work will only use DBS data as a demonstration of the ability of the diagnostic to forecast ELMs but we acknowledge that a burning plasma relevant multi-diagnostic approach~\cite{rothstein2026enabling} for such tasks is more robust and will likely be a direction for future work.}  

We previously applied convolutional neural networks to post hoc detection of ELMs from DBS data~\cite{teo2024using}. This work builds on that study and presents a proof-of-concept demonstration of using neural networks and DBS data to forecast the first ELM crash after the L-H transition. We trained a model on DBS data from multiple DIII-D discharges, tasked with using \msSI{50} of DBS spectrogram data to predict the probability of an ELM occurring within time windows of \msSI{0} to \msSI{50}, \msSI{50} to \msSI{100}, \msSI{100} to \msSI{150}, and beyond \msSI{150}. Similar work was previously conducted using beam emission spectroscopy (BES); a neural network was successfully trained to predict the probability of ELM crashes \cite{joung2024tokamak}. However, BES is an optical technique and is also reliant on neutral beam injection (NBI) \rev{and} is not planned for next-generation devices such as SPARC and STEP~\cite{creely2023sparc, wilson2025heating}. Ultimately, we seek to build tools for real-time forecasting ELMs in the burning plasmas of the next generation of tokamaks, with which DBS will likely be compatible. We present this effort in the rest of the paper. Section~\ref{sec:methods} briefly explains the model and neural network we have chosen for our forecasting task, and describes the data processing, training, and testing procedures of this project. Section~\ref{sec:results} evaluates the results of our model and discusses future work. Section~\ref{sec:conclusion} summarizes our work.

\section{Methods\label{sec:methods}}
 \subsection{Model}
The nature of our problem closely resembles that of survival analysis. Survival analysis originated from an interest in predicting a time-to-event, based on historical data. In the medical field, this time-to-event is generally the total time from diagnosis to death of a patient. It provides a framework that takes certain input features---such as patient weight---and, with a chosen model, provides an estimate of patient survival probability over time. We adapt a survival analysis model called DeepHit~\cite{lee2018deephit} to forecast the first ELM crash in DBS data. Traditional survival analysis models, such as the Cox proportional hazards model, are semi-parametric and analytic in nature. They estimate a risk score representing the relative hazard. which is a proxy for the likelihood of an event occurring at a given time, conditional on survival up to that point. The DeepHit model aims to directly predict the probability distribution of the time-to-event by discretizing time into bins and using a neural network to produce a probability mass function (PMF) of the time-to-event. This data driven approach is capable of modeling non-linear relationships between input features. In a model, the loss function is a minimized quantity to optimize the neural network during training; it is a quantity that reflects the performance of the network in completing a given task. The DeepHit loss function comprises of two parts,
\begin{equation}
    L=\alpha L_1+\beta L_2,
\end{equation}
where $\alpha$ and $\beta$ are loss scalers.
$L_1$ is a likelihood loss given by,
\begin{equation}
    L_1=-\sum\limits^N_{i=1}\log{P\left(T=\tau_{i}|X_i\right)},
\end{equation}
where $L$ is the loss function, $N$ is the total number of samples, $i$ is a particular sample, $T$ is a random variable representing the time-to-event, $\tau$ is the actual time bin when the event occurred, and $X$ is the input covariates to the neural network. The likelihood loss penalizes the neural network when the predicted probability of the event occurring at the actual time of event is low. $L_2$ is a ranking loss given by,
\begin{equation}
    L_2=\sum^N_{i=1}\sum^N_{j=1 \atop t_i<t_j}\phi\left(S(t_i|X_j)-S(t_i|X_i)\right).
\end{equation}
Here, $\phi$ is a convex loss function defined by $\phi(x)=\exp{\left(-\frac{x}{\sigma}\right)}$, where $\sigma$ is a scaling parameter. $S$ is the survival function defined by $S(t)=P(T>t)$, which is the probability that the time-to-event, $T$, is greater than a given time-to-event, $t$. The ranking loss penalizes the neural network when pairs of samples are ordered incorrectly. It should be noted that these loss functions are simplified from the original loss function from the DeepHit model as our problem does not involve multiple categories of events or censored data. For our model, the time-to-event is the time to the first ELM, and is discretized into four bins: \msSI{0} to \msSI{50}, \msSI{50} to \msSI{100}, \msSI{100} to \msSI{150}, and \msSI{150} to $\infty\mskip3mu\mathrm{ms}$. \rev{The $\alpha$ and $\beta$ loss scalers were set to 10 and 5, respectively, for our model. The results were not overly sensitive to these values, and as such minimal hyperparameter tuning was performed. We set a larger weight on $L_1$ to prioritize accuracy over ordering, which produced slightly better results.}

DeepHit is a framework used to train and utilize a neural network for forecasting and is independent of the neural network; the DeepHit framework can be applied to any neural network architecture. We opted not to adopt the original neural network architecture proposed by DeepHit, as the characteristics of our input data differ significantly from those typically encountered in survival analysis tasks. Instead, our neural network architecture largely follows that of the ResNetTransformer~\cite{zhang2021rest} (ResT), with only the final layer altered to output a PMF. ResT builds upon the well established convolutional neural network, ResNet~\cite{he2016resnet}, and integrates attention layers. In ResT, convolutional blocks with residual connections extract hierarchical local representations, which are then refined by transformer encoder blocks that model global feature interactions across the spatial dimensions. \rev{Further details of the architecture \revv{are provided in} \textcite{zhang2021rest}.} It is traditionally used to process image data, and can be applied to image classification tasks by training the model on a labeled dataset of images. In this project, ResT is utilized to process spectrogram data to capture short-range and long-range dependencies in frequency and time. The network forecasts an event by taking a DBS spectrogram segment containing past information as input and producing a PMF of the time to event. \rev{We designed our network to be similar to the smallest ResT model, ResT-lite, to minimize computational cost~\cite{zhang2021rest}. The ResT backbone consists of four stages with embedding dimensions of 64, 128, 256, and 512. The corresponding numbers of attention heads are 1, 2, 4, and 8, respectively. The transformer depths are 1, 2, 6, and 2 blocks per stage, while the spatial reduction (SR) ratios used in the attention layers are 8, 4, 2, and 1. This configuration results in a neural network with \SI{13.3}{M} learnable parameters.}

\begin{figure}
    \centering\includegraphics[width=\linewidth]{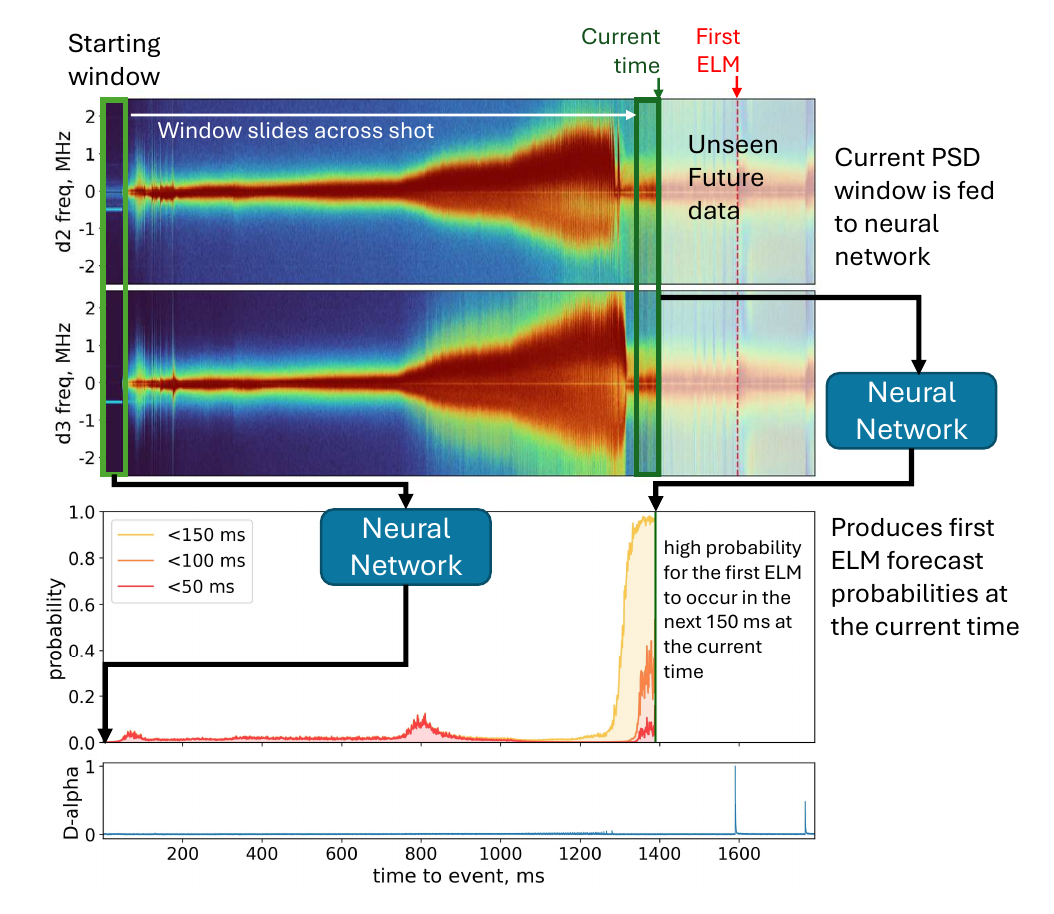}
    \caption{\rev{Illustration of the implementation of the model for ELM forecasting during an experiment.} \revv{The bottom signal is the normalized divertor D$_\alpha$ light intensity (a.u.). The L-H transition occurs at \msSI{1289} and the first ELM occurs at \msSI{1589}.} At the start of a shot, the model takes as input a window (green box) spanning \msSI{50} of DBS spectrogram from channels 2 and 3 \revv{shown at the top} and produces predicted probabilities of the first ELM occurring within the next \msSI{150}, \msSI{100}, and \msSI{50} measured from the end of the input window. As the plasma evolves, updated spectrogram windows are fed to the model at a set interval in real time to obtain forecasts of the aforementioned probabilities. The \revv{middle} graph shows the time series of these probabilities with each data point representing the output of the model on a spectrogram window. The most recent probability estimate serves as a warning of a potential ELM crash. To mimic this, we test our model by sliding the window across a prerecorded shot in \msSI{1} steps, measuring how the output probabilities change with time.}
    \label{fig:model_scheme}
\end{figure}

A schematic of how the model functions is shown in figure~\ref{fig:model_scheme}. \msSI{50} of DBS spectrogram data is given to the neural network as input, which we refer to as a window. Following the DeepHit framework, the network outputs four distinct probabilities. Each probability represents the predicted probability of an ELM crash occurring within one of the four predefined time bins. For easier interpretation, these binned probabilities can be transformed into the probability of the first ELM occurring within the next \msSI{150}, \msSI{100}, and \msSI{50} from the time at the end of the time window. In an operational experiment, these spectrogram windows will be fed to the model at set intervals in real time to obtain updated estimates of the forecast probabilities. The probability estimates \revv{serve} as a warning for a possible ELM crash.

\subsection{Data}\label{sec:data}
\begin{figure}
    \centering
    \includegraphics[width=0.9\linewidth]{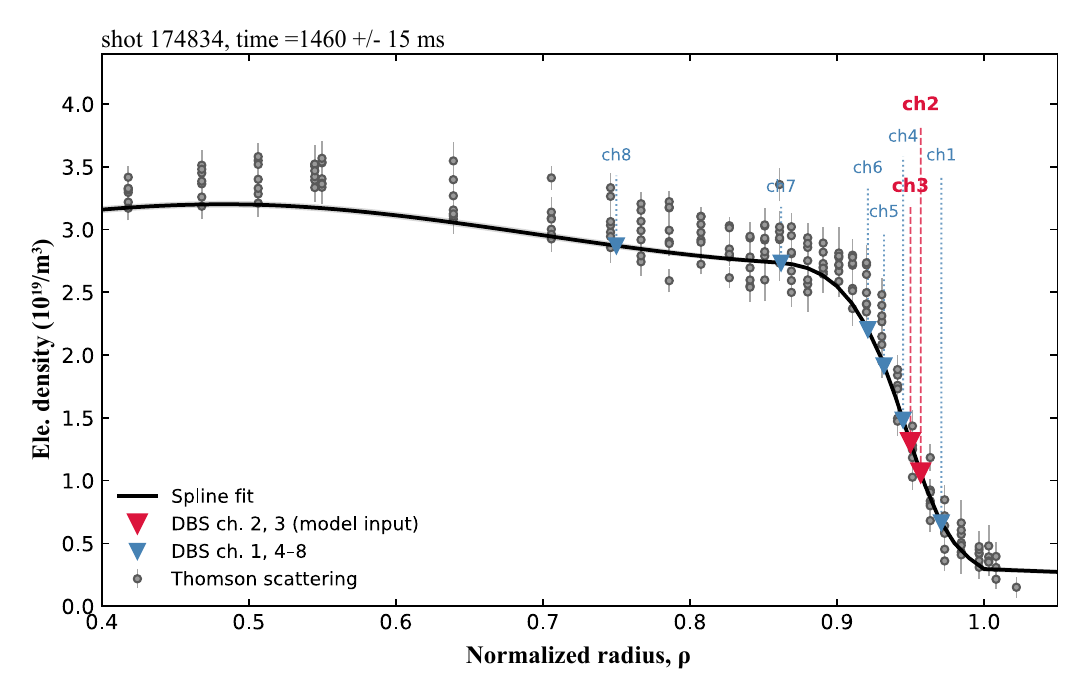}
    \caption{\rev{Electron density profile for shot 174834 at $t=\msSI{1460}$, shortly before the first ELM. Gray points: Thomson scattering measurements; black curve: spline fit with uncertainty band. Triangles mark the cutoff locations of the eight DBS channels from GENRAY ray tracing; channels 2 and 3 (red), localized in the steep-gradient region of the pedestal, provide the spectrogram input to the forecasting model.}}
    \label{fig:profile}
\end{figure}
All data used in this research were obtained from the DIII-D database~\cite{Buttery:DIIID:2023}. The DIII-D experiment is the largest operating tokamak in North America that pioneers research in fusion energy and plasma physics. \rev{This study includes \revv{26} different neutral-beam-heated discharges\revv{---16 for training, 4 for validation and 6 for testing---}with input power close to the L-H power threshold, which normally develop large type-I ELMs of the kind considered most damaging to the divertor and plasma-facing components. They are representative of the reference discharge that DIII-D runs each morning before operations to verify machine reproducibility, which is why a relatively homogeneous set of this discharge type is available.} 
The input data consists of \msSI{50} spectrogram segments from two DBS channels---2 and 3, corresponding to probe beams of \SI{57.5}{\giga\hertz} and \SI{60}{\giga\hertz}, respectively. As ELM onset is governed by pedestal dynamics, we chose the channels that have cutoffs in this region \rev{as shown in figure~\ref{fig:profile}}. The measurement locations of channels 2 and 3 are localized to the steep gradient region of the pedestal at $\rho\simeq0.961$ and $\rho\simeq0.955$, respectively. To estimate the radial locations of the channels, we used experimental profiles and plasma equilibrium as an input to GENRAY raytracing calculations~\cite{smirnov2001genray}. For each channel, we first generate spectrograms of the spectral power by extracting \msSI{50} segments from the DBS output waveforms. \rev{The spectrograms are computed from the complex quadrature signal I + iQ. The DBS receiver records the in-phase (I) and quadrature (Q) components, 90° apart, which together contain both the amplitude and the phase of the backscattered field; the two-sided quadrature spectrum therefore retains the sign of the Doppler shift and hence the flow information, in addition to the fluctuation amplitude.} The waveforms were standardized to a sampling frequency of \SI{5}{\mega\hertz}; data that was collected at \SI{8}{\mega\hertz} was downsampled with a lowpass filter and polyphase resampling with \texttt{scipy.signal}. After which we used \texttt{scipy.signal} to obtain the power spectrum density (PSD), using a Hann window and a segment length of 512 time-points. The resulting spectrogram was transformed to logarithmic space in power to compress the dynamic range. \rev{Additionally, we downsampled the spectrogram to 256 by 256 using \texttt{skimage.transform} to reduce dimensionality due to hardware limitations.}  
\rev{This work focuses only on forecasting the first ELM after L-H transition, and as such the DBS data used for training and testing only consists of times before the first ELM event.} The first ELM was detected using filterscopes looking at the divertor region and measuring D$_\alpha$ intensity, which monitor the \SI{656.1}{nm} Balmer-alpha emission peaking that happens during an ELM crash. The initial ELM following the L-H transition was identified using the ELM module in OMFIT~\cite{meneghini2013integrated, meneghini2015integrated}. The time-to-event for a particular spectrogram can then be calculated by taking the difference between the time of the first ELM and the last recorded time of the spectrogram. The time-to-event values were discretized into time bins to produce the array representation used as the ground truth for the DeepHit model.

\subsection{Training and Testing}
\begin{figure}
    \centering
    \includegraphics[width=0.5\linewidth]{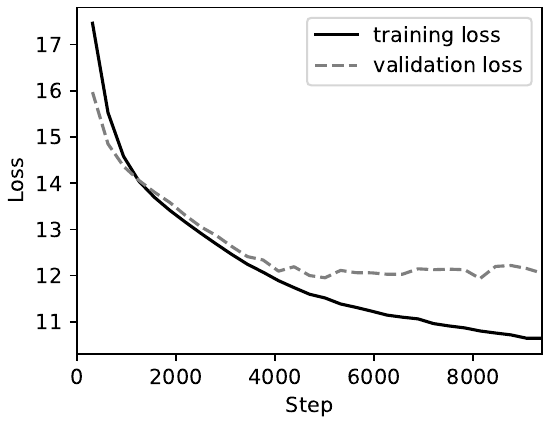}
    \caption{\rev{Training and validation loss with training time. The y-axis shows the DeepHit training loss and the x-axis shows the number of training steps (312 steps correspond to 1 epoch).}}
    \label{fig:logs}
\end{figure}
Training and testing of the model were implemented using the Python PyTorch package. The data processing and machine learning code are available on GitHub, see section below on data availability.

We sampled \num{4e4} spectrograms from 16 shots---174817, 174822, 174823, 174825 to 174827, 174829, 174830, 174833, 184437, 184438, 184483, 184485, 184488, 184480, and 184481---for training and \num{1e4} \revv{spectrograms} from 4 shots---174819, 174831, 184439, and 184482---for validation. We chose to sample the validation dataset from a separate selection of shots to prevent data leakage from overlapping sections of spectrograms within a particular shot. These shots contain the L-H transition and type-I ELMs and have been previously studied for inter-ELM electron density fluctuation behavior~\cite{barada2021new}. Sampling involved clipping (trimming to a specific range) the waveforms from each shot from the start of the DBS recording until the first ELM crash. From the clipped data, we sampled \msSI{50} segments. To avoid class imbalance, each time-to-event bin is allocated an equal number of samples by grouping all possible samples by their respective bin and randomly sampling a set number of samples from each of these groups. If there were insufficient samples in a specific bin, all samples in that bin were used. These segments were processed to produce spectrograms as described in section \ref{sec:data} and used as inputs for the model. A discretized time-to-event array was produced for each spectrogram to serve as the ground truth. In neural network training, the optimizer is an algorithm for the gradient descent optimization process, and the scheduler determines how the learning rate, the gradient descent step size parameter, changes with training time. We used the \texttt{AdamW} optimizer with a learning rate of \num{3e-7} and weight decay of \num{1e-2}, and \texttt{ExponentialLR} scheduler with a gamma value of 0.99. \rev{These values were chosen after hyperparameter tuning to ensure the model training was stable. We used a batch size of 128, which we found was a good compromise between training stability and speed.} The model was trained for 30 epochs on an Nvidia A4000 GPU. Each epoch represents a single pass of network training over the entire dataset. \rev{The testing loss stabilized and plateaued after epoch 15; epoch 25 had the lowest testing loss and thus was used for testing. We show the training and validation loss curves in figure~\ref{fig:logs}.}

We tested on 6 shots---174818, 174832, 174834, 184440, 184486, and 184489. We selected these shots to ensure variation in the time elapsed between the L-H transition and the first ELM, enabling assessment of the model’s ability to generalize. Each shot is clipped following the same procedure applied to the training data. \msSI{50} segments are then sampled at intervals of \SI{1}{\milli\second}. The spectrograms were fed through the model to produce PMFs. The temporal order of the spectrograms was preserved to evaluate model performance from the start of the shot to the first ELM event. This approach reflects a real-time forecasting scenario where the model is fed updated DBS data at a cadence of \SI{1}{\milli\second}, thereby providing an updated forecast of the first ELM at that frequency (figure~\ref{fig:model_scheme}).

\section{Results and Discussion\label{sec:results}}
\subsection{Model Results}
We evaluate the model based on how it would function as an alert for ELMs in a control system. For each PMF output of the model, we can calculate the predicted probability of an ELM crash within the next $t\,\mathrm{ms}$ with the following formula,
\begin{equation}
    P(T<t)=\sum\limits_{i\atop b_i<t}p_i,
\end{equation}
where $p_i$ is the probability mass of bin $i$, and $b_i$ is the upper limit of bin $i$. For our model, we set $t$ to be \msSI{150}, \msSI{100}, and \msSI{50}, each representing an alert level for an ELM crash. We track these alert levels with time for each shot to evaluate the performance of our model. We consider an alert to be triggered when the alert level crosses a probability of 0.5 for \msSI{5} consecutively. \rev{This was chosen heuristically and not optimized to the testing shots so as to avoid artificial inflation of the model performance.}

\begin{table}[]
    \centering
    \begin{tabular}{lcc} \toprule
         \textbf{Alert} & \textbf{AUC} & \textbf{Brier} \\ \midrule
         \phantom{1}\msSI{50} & 0.92 & 0.07 \\ 
         \msSI{100} & 0.95 & 0.09 \\
         \msSI{150} & 0.70 & 0.27 \\ \bottomrule 
    \end{tabular}
    \caption{\rev{Performance of the model is evaluated using the area under the curve (AUC) and the Brier score. The AUC measures the model’s ability to correctly rank samples by risk, with a value of 1.0 indicating perfect ordering. The Brier score measures the accuracy of the predicted probabilities via mean squared error, where 0.0 corresponds to perfect probabilistic calibration. Note that to avoid inflating performance from early time steps where the task is trivial, \revv{these metrics are computed only for spectrogram windows from the final 100 ms preceding the L-H transition to the first ELM.}}}
    \label{tab:segment_metrics}
\end{table}

\begin{figure}
    \centering
    \includegraphics[width=0.45\linewidth]{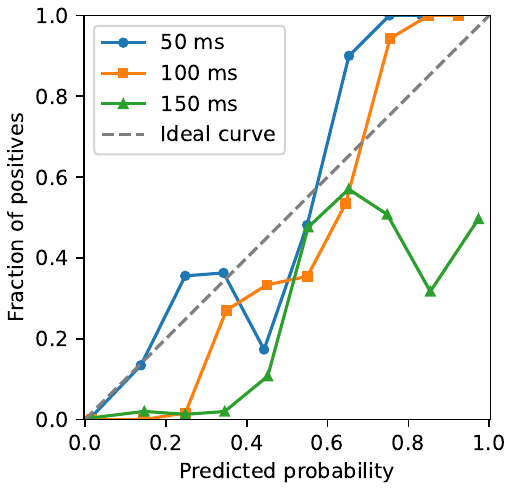}
    \caption{\rev{Reliability curve for each of the alert levels which compares predicted probabilities to observed outcome frequencies across probability bins. Only the \msSI{100} alert shows a common monotonic calibration pattern with overconfidence at low predicted probabilities and underconfidence at high predicted probabilities. The other alert levels are non-monotonic with local miscalibrations.}}
    \label{fig:reliability}
\end{figure}

\rev{Table~\ref{tab:segment_metrics} shows the model performance with the area under the curve (AUC) and the Brier score. \revv{These metrics evaluate the model predictions for individual spectrogram windows by comparing the predicted risk and probabilities with the corresponding ground truth.} The AUC measures the model’s ability to correctly rank samples by risk, with a value of 1.0 indicating perfect ordering. The Brier score measures the accuracy of the predicted probabilities via mean squared error, where 0.0 corresponds to perfect probabilistic calibration. To avoid inflating performance from early time steps where the task is trivial, \revv{these metrics are computed only for windows from the final 100 ms preceding the L-H transition to the first ELM.} This evaluation window focuses the assessment on the period where the predictive signal emerges and is most relevant for early warning. The \msSI{100} and \msSI{50} perform well in both metrics while the \msSI{150} alert performs significantly worse. This is attributed to the fact that the model learned to trigger this alert immediately after the L-H transition and not \msSI{150} before the first ELM. This will be discussed later in more detail. Figure~\ref{fig:reliability} shows the reliability curve for each of the alert levels. The curve compares predicted probabilities to observed outcome frequencies across probability bins to assess how well-calibrated a model’s probabilistic outputs are. The \msSI{100} alert level shows a common monotonic calibration pattern with  overconfidence at low predicted probabilities and underconfidence at high predicted probabilities. The other alerts are not well calibrated as probabilities, with non-monotonic behavior and larger deviation from the ideal curve.}

\begin{figure}[!t]
    \centering
    \begin{tabular}{cc}
        \begin{subfigure}[t]{0.45\textwidth}
            \centering
            \includegraphics[width=\linewidth]{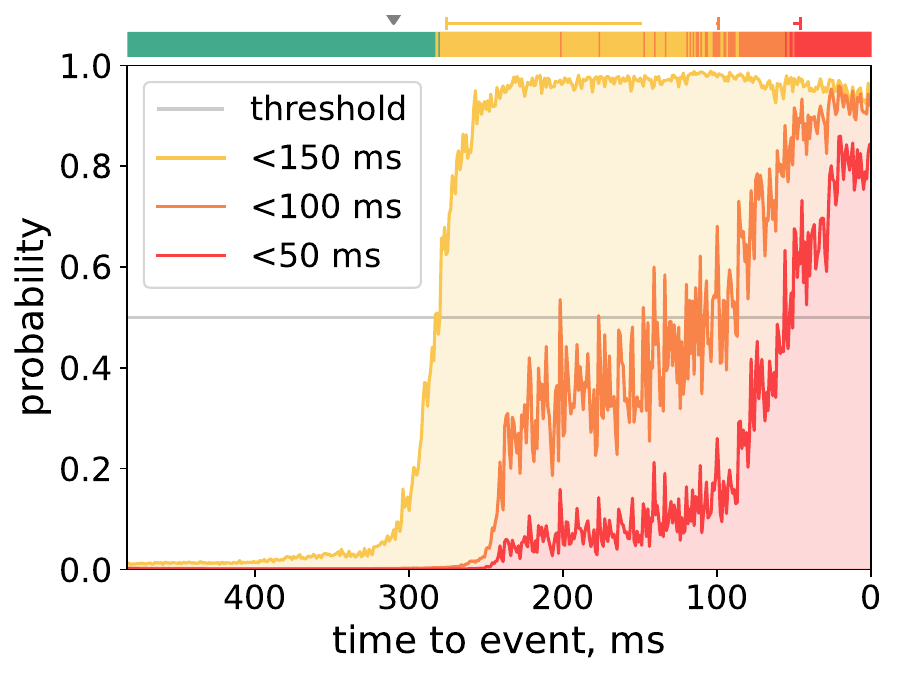}
            \caption{Shot 174818\label{fig:174818}}
        \end{subfigure}
        &
        \begin{subfigure}[t]{0.45\textwidth}
            \centering
            \includegraphics[width=\textwidth]{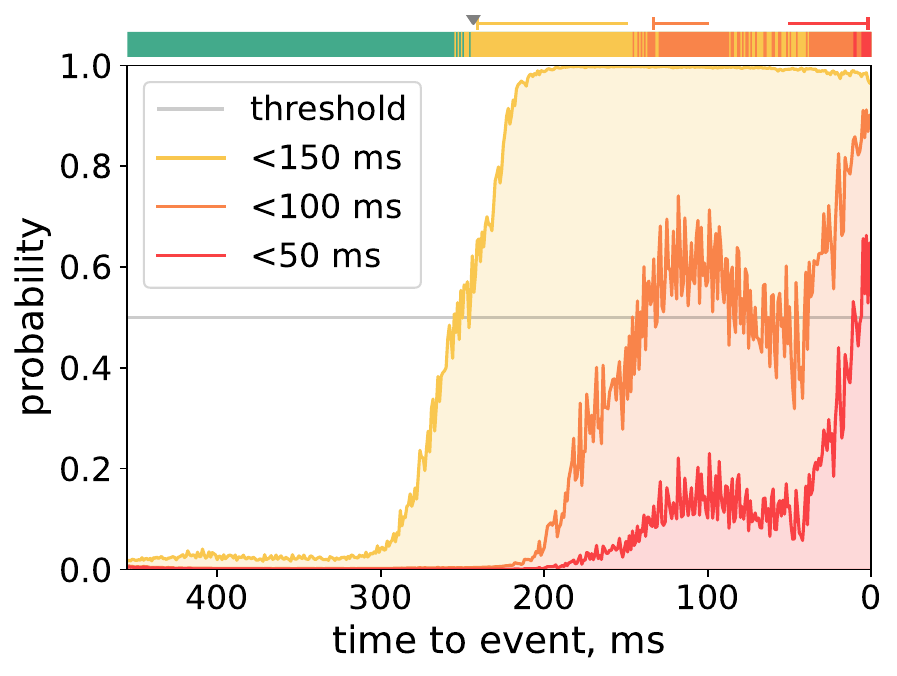}
            \caption{Shot 174834\label{fig:174834}}
        \vspace{.5cm}
        \end{subfigure}\\
        \begin{subfigure}[t]{0.45\textwidth}
            \centering
            \includegraphics[width=\textwidth]{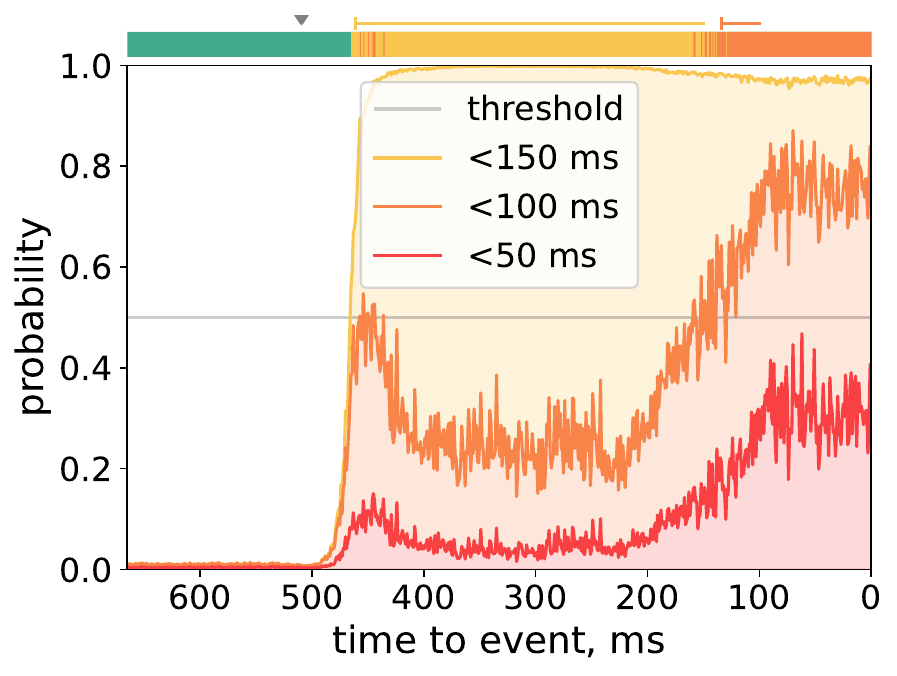}
            \caption{Shot 184486\label{fig:184486}}
        \end{subfigure}
        &
        \begin{subfigure}[t]{0.45\textwidth}
            \centering
            \includegraphics[width=\textwidth]{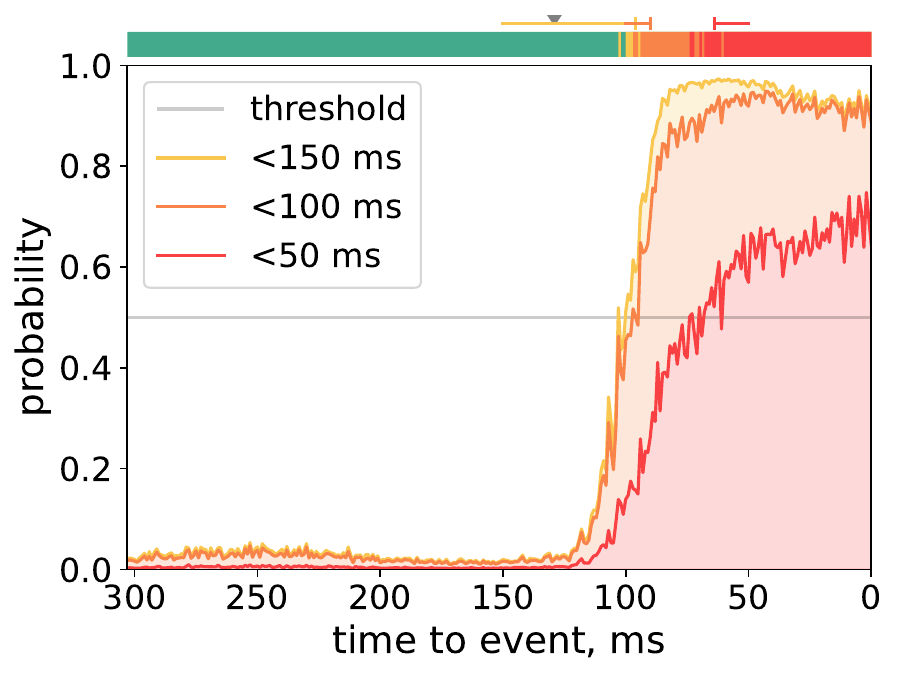}
            \caption{Shot 184489\label{fig:184489}}
        \end{subfigure}
    \vspace{1cm}    
    \end{tabular}
\caption{Model results of four shots, each plot shows three alert levels---\msSI{150} in yellow, \msSI{100} in orange, and \msSI{50} in red. \rev{The x-axis shows the time before the first ELM where \msSI{0} is when the first ELM occurs.} The alert is considered to be triggered once the alert level remains above a threshold of 0.5 for \msSI{5} consecutively. Above each plot, the horizontal bar represents the highest alert that has crossed the 0.5 threshold, with green representing no alerts crossed the threshold. The timestamps when the alert triggers are indicated with the vertical line marker with the horizontal line extending to the respective \revv{reference} time (the time-to-event that the alert is trained to trigger on). The gray arrow represents the end of the L-H transition when H-mode begins.}
\label{fig:results}
\end{figure}

\begin{figure}
    \centering
    \includegraphics[width=0.7\linewidth]{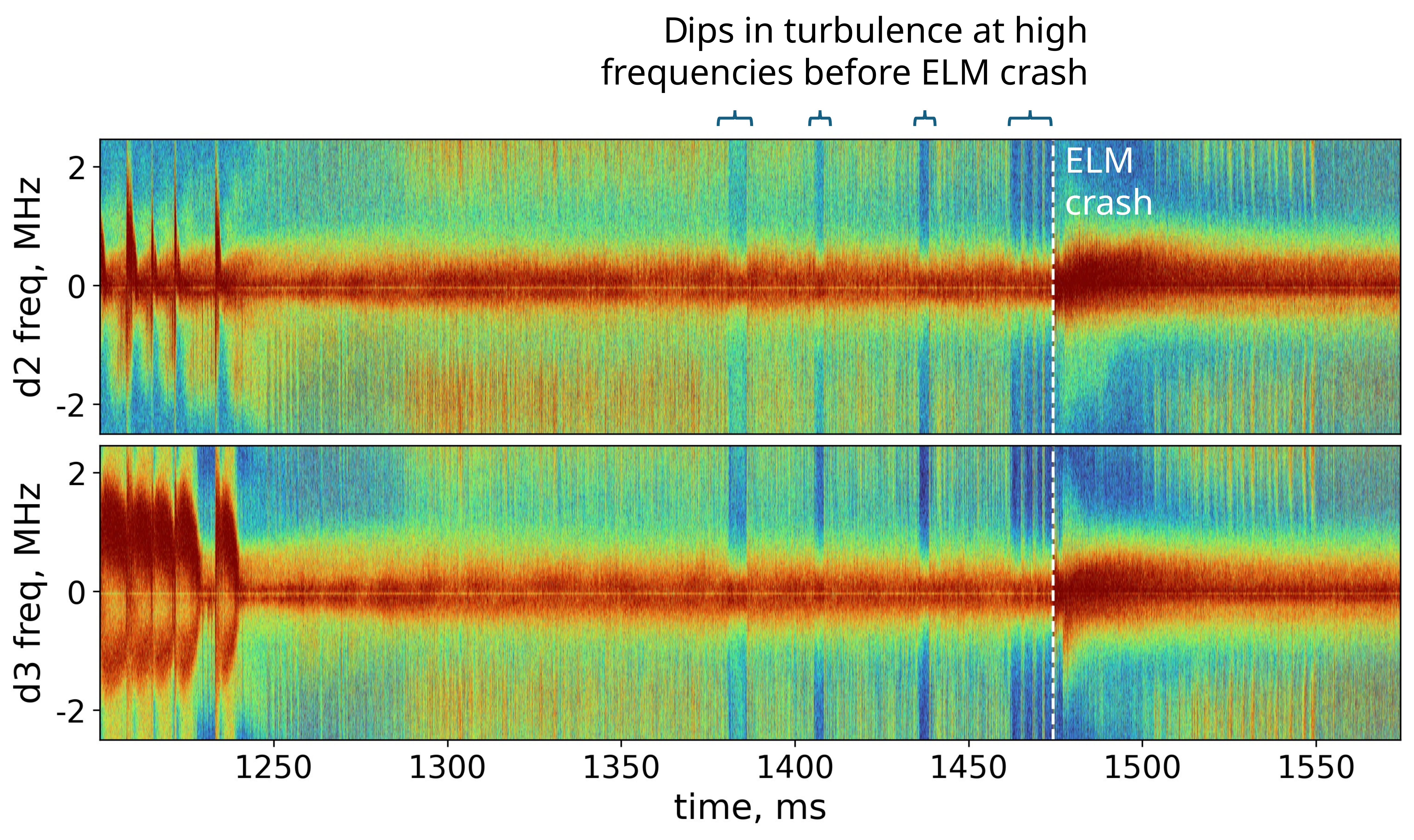}
    \caption{PSD of shot 174834 from the \revv{L-H transition} to the first ELM (white dotted line). The turbulence at high frequencies dip sporadically before the ELM crash occurs.}
    \label{fig:174834data}
\end{figure}

\begin{figure}
    \centering
    \includegraphics[width=0.5\linewidth]{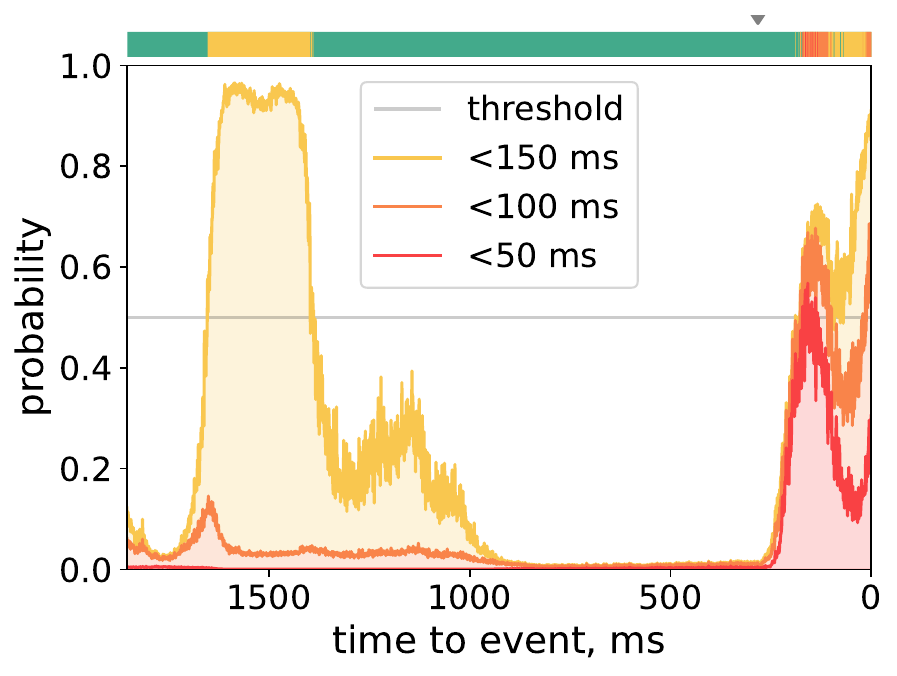}
    \caption{Model results for shot 184440. \rev{The x-axis shows the time before the first ELM where \msSI{0} is when the first ELM occurs.} There are three alert levels---\msSI{150} in yellow, \msSI{100} in orange, and \msSI{50} in red. The alert is considered to be triggered once the alert level remains above a threshold of 0.5 for \msSI{5} consecutively. Above each plot, the horizontal bar represents the highest alert that has crossed the 0.5 threshold, with green representing no alerts crossed the threshold. The gray arrow represents the end of the L-H transition when H-mode begins.}
    \label{fig:184440}
\end{figure}

Figure~\ref{fig:results} show our model results from shots 174818, 174834, 184486, and 184489. We show only the results of the last \msSI{300} to \msSI{700} as the alert levels before these windows remain consistently below 0.1 and do not contain any interesting insights. Above each plot, we show the highest alert that has crossed the 0.5 threshold as a visualization of the alert level at a given time. Green, yellow, orange, and red represent no alerts, the \msSI{150} alert, the \msSI{100} alert, and the \msSI{50} alert, respectively. The results of shot 174818 are shown in figure~\ref{fig:174818}. We observe the \msSI{150} alert peaking early at \msSI{276}. Checking the DBS spectrogram, we found that this timing coincides with the L-H transition. Essentially, the model here learned to identify H-mode before the crash although the model is not trained specifically to detect L-H transition. While unexpected, this result is logical as ELMs occur only after L-H transition.  As will be seen below, this behavior remains consistent in the results of other shots. The \msSI{100} alert sees an initial sharp increase about \msSI{50} after the L-H transition before plateauing around the 0.4 to 0.5 mark. At the \msSI{100} mark, the alert level sharply increases again. \revv{The initial increase may occur when the L-H transition moves out of the 50 ms spectrogram window, such that the entire input window consists of H-mode data.} The \msSI{100} alert triggers at \msSI{99}. The \msSI{50} alert displays behavior similar to that of the \msSI{100} alert, with an initial sharp increase, followed by a plateau and a final sharp increase again. The trigger for the \msSI{50} alert occurs at \msSI{46}. The results of shot 174834 are shown in figure~\ref{fig:174834}. The \msSI{150} alert triggers at \msSI{241}, where the L-H transition occurs. The \msSI{100} alert increases sharply \msSI{50} after the L-H transition, but triggers the alert early at \msSI{133} before dropping slightly and increasing sharply again \msSI{40} before the first ELM, producing a false peak. Inspecting the DBS spectrogram reveals large dips in turbulence in the high frequencies where the alert level first peaks shown in figure~\ref{fig:174834data}. \rev{We believe the model responds to the evolution of the broadband turbulence amplitude as the pedestal builds, including the intermittent dips in the high-frequency part of the spectrum that occur before the first ELM. We would characterize these dips as a feature of the evolving pedestal state — reflecting the coupled dynamics of the pedestal gradients, E×B shear, and turbulence — rather than as an ELM precursor in the usual sense, since they are not observed before subsequent ELMs.} The \msSI{50} alert has a similar response to the \msSI{100} alert but does not result in an early trigger. Importantly, the alert level remains low until \msSI{40} to the first ELM and only triggers at \msSI{2}. The results of shot 184486 are shown in figure~\ref{fig:184486}. The \msSI{150} alert occurs at \msSI{461} where the L-H transition occurs. The \msSI{100} alert increases sharply with the \msSI{150} alert in this case but peaks around 0.4 to 0.5 marks and stays around that level until about \msSI{200} to the first ELM and triggers at \msSI{134}. The \msSI{50} alert shares similar behavior with the \msSI{100} alert, increasing with time. However, for this shot, the alert is never triggered, only peaking around 0.4. The results of shot 184489 are shown in figure~\ref{fig:184489}. Again, the \msSI{150} alert triggers when the L-H transition occurs, at \msSI{96}. We observe the \msSI{100} alert triggering at \msSI{90}, largely following the curve of the \msSI{150} alert. The \msSI{50} alert started increasing at the same point as the other alerts but with a smaller gradient, triggering at \msSI{64}. The model here was able to recognize correctly that the ELM occurs quickly after the L-H transition. Testing on shot 184440 as seen in figure~\ref{fig:184440}, we observe high frequency noise in the L-mode regime which was incorrectly detected as H-mode by the model. In H-mode, we observe that the \msSI{100} alert \revv{exhibits} a false peak, triggering the alert early at \msSI{167}. It falls below the 0.5 level, only increasing about \msSI{50} to the first ELM. The \msSI{50} alert level did not trigger in this shot. This result is expected since the shot is dissimilar to the training data due to differing plasma conditions.

\begin{table}[]
    \centering
    \begin{tabular}{lcrrcccc} \toprule
        \textbf{Alert} & \textbf{Ref. time} & \textbf{Mean} & \textbf{Median} & $\boldsymbol{\pm}\textbf{10}\,\textbf{ms}$ & $\boldsymbol{\pm}\textbf{25}\,\textbf{ms}$ & \textbf{FN} & \textbf{FP} \\ \midrule
        \phantom{1}\msSI{50} & \phantom{1}\msSI{50} & \msSI{-20.7} & \msSI{-24.5} & 1 & 2 & 2 & 0 \\ 
        \msSI{100} & \msSI{100} & \msSI{+20.5} & \msSI{16.5} & 3 & 3 & 0 & 0 \\ 
        \msSI{150} & \msSI{150} & \msSI{333.7} & \msSI{108.5} & 0 & 0 & 0 & 3 \\ 
        \msSI{150} & L-H transition & \msSI{207.1} & \msSI{-19.6} & 2 & 0 & 0 & 1 \\ \bottomrule
    \end{tabular}
    \caption{\rev{\revv{Performance of the model with respect to alert trigger times.} The tables shows the mean and median times of the difference between the alert trigger times and the reference time (ref. time), with positive values indicating an early trigger and negative values indicating a late trigger. Also shown is the number of times, out of the six testing shots, the model predicting within \msSI{10} ($\pm$\msSI{10}) and \msSI{25} ($\pm$\msSI{25}) of the reference time. Lastly, the table shows the number of false negatives (when the alert did not trigger) and false positives (when the alert triggered more than \msSI{100} before the reference time).}}
    \label{tab:time_metrics}
\end{table}

\rev{Additionally, we can evaluate the model performance with respect to the trigger alert times. Table~\ref{tab:time_metrics} provides a summary of the results. The first column shows the alert level and the second column shows the reference time that we evaluate this alert level on. The next two columns show the mean and median of the difference between the alert trigger time and the reference time. For the \msSI{50} alert we observe a tendency to trigger late and for the \msSI{100} alert a tendency to trigger early. The \msSI{150} alert performs very poorly when compared to a reference time of \msSI{150}, but much better when compared to the L-H transition, consistent with the results above. The large mean values here are due to the early trigger in shot 184440 skewing the result; the median here is a better representation of the general performance. The next two columns show, out of the six testing shots, the number of times the model predicts within \msSI{10} ($\pm$\msSI{10}) and \msSI{25} ($\pm$\msSI{25}) of the reference time. The \msSI{100} alert performs the best here with the alert triggering within \msSI{10} for three shots. The remaining three shots triggered outside $\pm\msSI{25}$, with all triggering early rather than late. The last two columns show the false negatives and false positives, with false negatives defined as the alert not triggering before the first ELM and false positives as the alert triggering \msSI{100} before the reference time. As consistent with the tendency to trigger late, the \msSI{50} did not trigger in two of the shots. The \msSI{150} alert, with a L-H transition as the reference time, has one false positive. This early trigger is from shot 184440 as discussed early.} 

These preliminary results of our model are favorable. While the \msSI{150} alert has very low accuracy predicting the first ELM \msSI{150} before crash, we deduced that the model was detecting H-mode, and performs this task with high accuracy. We have, albeit unintentionally, demonstrated the ability of a machine learning model to measure H-mode from DBS data. While it is not the intended result of our model, this result may have some value, since traditional optical methods for measuring the \revv{H-mode} in plasmas may not be ideal for future tokamaks~\cite{creely2023sparc, chrystal2020predicting}. The \msSI{100} alert is fairly consistent in accurately triggering \msSI{100} before the first ELM, even with variability in duration between the L-H transition and the first ELM. This result is extremely promising, as it showcases the ability of the model to trigger an alert sufficiently early to activate ELM suppression systems. Current RMP systems can activate within \msSI{50}~\cite{mckee2013increase}, providing a good margin to work with the forecasting model. However, it is important to note that suppression requires a variable amount of time after RMP activation, depending on the plasma parameters, ranging from tens to hundreds of ms. Thus, the current forecasting model may not prevent all ELMs from occurring but will allow for RMP activation well before the first ELM. On the other hand, the \msSI{50} alert does not display the same consistency, triggering late or not at all. \rev{Another recent approach that has been applied successfully in KSTAR tokamak~\cite{kim2023integrated} is the feed-forward application of RMP with respect to the L-H transition time to avoid the first ELM. However, even from a 26 shots distribution that we have used for this work, there is a variation of 100-700 ms in the time window from L-H transition to the first ELM. A forecasting model like we have developed here may work better because this model learns the signatures of turbulence evolution leading up to the first ELM.} 

\rev{On an NVIDIA L40S GPU, the model can run at \SI{226}{\hertz} or \msSI{4.5} per spectrogram if each spectrogram is fed sequentially in batches of 1. With the current model, achieving the \SI{1}{\kilo\hertz} frequency in the testing scenario demonstrated in this study is possible if spectrograms are processed in batches of 10, where the model frequency increases to \SI{1.8}{\kilo\hertz} or \msSI{0.56} per spectrogram. Alternatively, multi-GPU techniques or model optimization and sparsification can be applied to increase the model frequency.}

\subsection{Future Work}
\rev{To progress this work beyond a proof-of-concept, judicious optimization of the model should be performed. There are various aspects, such as model size, network architecture, hyperparameters, and trigger conditions that require deeper analysis.} To improve model generality, especially since discharges across experiments display significant variability, an important step is to identify which types of discharges will be relevant for training and add these shots to the training dataset. The model may also benefit from training on more types of data. Currently, only the DBS spectrogram is used as input, but the model can be designed to process additional data such as from other mm-wave diagnostics or experimental parameters. Due to the limited size of the dataset used for training and testing, it is difficult to reliably quantify the model performance. Future work will involve collecting more shots for testing to obtain distributions of the alert trigger times. The mean and spread of these distributions will provide key insights for model performance analysis and consequently inform strategies to improve the model. \revv{Since the starting ramp phase is not relevant for ELM forecast and occasionally contain noise, only using data \msSI{500} before the L-H transition may improve training stability.} \rev{Additionally, we plan to apply attention-map interpretability methods to infer physical relationships encoded by the model, potentially revealing new insights into the physics governing plasma behavior.} A further area of work is to develop a model that can be integrated into the plasma control system or PCS of DIII-D for real-time forecasting and ELM mitigation/avoidance. To achieve this, the model has to run sufficiently fast, and the software should be optimized for deployment. In this context, it may be valuable to explore conventional time-series machine learning approaches, such as long short-term memory networks and attention-based architectures~\cite{hochreiter1997long, wu2022timesnet}. The model may also be adapted to perform other forecasting tasks. An alternative application of the model would be forecasting disruptions~\cite{schuller1995disruptions, guo2021disruption, vega2022disruption} in tokamaks.

\section{Conclusion\label{sec:conclusion}}
In this preliminary study, we demonstrate the ability of adapting the DeepHit model and ResT framework to forecast first ELM crash after the L-H transition from DBS data. Our model is shown to be adept at this task. The \msSI{150} alert interestingly triggers at the L-H transition, essentially detecting H-mode. This may be useful to replace optical systems for detecting H-mode in \revv{tokamak} plasmas. The \msSI{100} alert performed well, triggering between \msSI{96} and \msSI{134} to the first ELM. This in an encouraging result as an alert provides sufficient time for RMP systems to activate. Future forecasting models and RMP systems will hopefully increase the margin. \rev{More broadly, the forecast probabilities our model outputs are actuator-agnostic---they could feed control logic for RMPs, ELM pacing, impurity/Li injection, or scenario adjustment.} On the other hand, the \msSI{50} alert displayed inconsistent behavior, tending to \revv{trigger} late with less than \msSI{10} to the first ELM in shots 174832 and 174834, and not triggering for shot 184486. This unreliability can hopefully be addressed with further training, improved models and better optimized alert trigger criteria. The immediate goals are to further optimize the model, improve model generality, and perform more rigorous testing. Ultimately, the aim is to develop a real-time forecasting tool for future tokamaks to provide early warnings for deleterious events.

%
%

\ack{ 
N.Q.X. Teo thanks the Nanyang Technological University (NTU) Singapore, CN Yang Scholars Program, for financial support.This research was partially supported by the FEAT SRTT, A*STAR and JST Moonshot R\&D Grant Number JPMJMS2011. This material is based upon work supported by the U.S. Department of Energy, Office of Science, Office of Fusion Energy Sciences, using the DIII-D National Fusion Facility, a DOE Office of Science user facility, under Award(s) DE-SC0024536, DE-SC0022563, DE-SC0019352, DE-FG02-08ER54984, and DE-FC02-04ER54698. Part of the data analysis was performed using the OMFIT integrated modeling framework~\cite{meneghini2015integrated, pratt2025doppler}.}

\section*{Disclaimer}
This report was prepared as an account of work sponsored by an agency of the United States Government. Neither the United States Government nor any agency thereof, nor any of their employees, makes any warranty, express or implied, or assumes any legal liability or responsibility for the accuracy, completeness, or usefulness of any information, apparatus, product, or process disclosed, or represents that its use would not infringe privately owned rights. Reference herein to any specific commercial product, process, or service by trade name, trademark, manufacturer, or otherwise does not necessarily constitute or imply its endorsement, recommendation, or favoring by the United States Government or any agency thereof. The views and opinions of authors expressed herein do not necessarily state or reflect those of the United States Government or any agency thereof.

\roles{N.Q.X. Teo (formal analysis, methodology, visualisation, original draft, review \& editing), K. Barada (data curation, review, conceptualization \& editing - supporting), V.H. Hall-Chen (project administration, funding acquisition, resources, supervision, and  resources, review \& editing - suporting, conceptualization - supporting), L. Gu (conceptualization  - supporting), T.L. Rhodes (data curation - supporting, funding acquisition).}

\data{The data that support the findings of this study are available from the corresponding author upon reasonable request. The repositories are available on Github through \url{https://github.com/NathanTeo/fusion_data_processing.git} and \url{https://github.com/NathanTeo/fusion_ml.git}}


\printbibliography

\end{document}